# Spatial coherence of random laser emission

Brandon Redding,[1,*] Michael A. Choma,[2] Hui Cao[1]

[1]*Department of Applied Physics, Yale University, New Haven, CT 06511*
[2]*School of Medicine, Yale University, New Haven, CT 06511*
*\*Corresponding author: brandon.redding@yale.edu*

We experimentally studied the spatial coherence of random laser emission from dye solutions containing nanoparticles. The spatial coherence, measured in a double-slit experiment, varied significantly with the density of scatterers and the size and shape of the excitation volume. A qualitative explanation is provided, illustrating the dramatic difference from the spatial coherence of a conventional laser. This work demonstrates that random lasers can be controlled to provide intense, spatially incoherent emission for applications in which spatial cross talk or speckle limit performance. © 2011 Optical Society of America

Over the past two decades, random lasers have been the subject of intense theoretical and experimental studies [1, 2]. Coherence is a fundamental characteristic of any laser, and, as such, the temporal coherence [3, 4] and second-order coherence [5-7] of random lasers have been thoroughly investigated. However, the spatial coherence of random laser emission is not well understood despite initial observations indicating that it is much lower than in a conventional laser [4, 8-10]. Not only is spatial coherence of fundamental interest, but since this characteristic is likely to be quite different for random lasers than for conventional lasers, it could lead to a host of applications in which random lasers could outperform conventional lasers. For example, optical coherence tomography [11] and laser ranging [12] are limited by spatial cross talk and speckle and could benefit from the development of an intense, spatially incoherent light source. To this end, we present a systematic, experimental investigation of the spatial coherence of random laser emission. Specifically, we consider the effect on spatial coherence of the scatterer concentration, excitation volume, and pump intensity. Based on this work, we are able to identify regimes of operation in which a random laser provides spatially incoherent emission which could be used as a speckle-free laser probe beam.

Our experiments were performed on a series of samples consisting of a laser dye solution and polystyrene spheres. The solution consisted of 5 mMol of Rhodamine 640 dissolved in diethylene glycol (DEG). The polystyrene spheres were each ~240 nm in diameter and their scattering cross section in DEG, $\sigma$, was calculated to be $1.67 \times 10^{-11}$ cm². We fabricated samples with polystyrene sphere concentrations, $\rho$, of $1.2 \times 10^{12}$ cm⁻³, $6.1 \times 10^{12}$ cm⁻³, and $1.2 \times 10^{13}$ cm⁻³. Since the average distance of adjacent scatterers was much larger than the diameter of the scattering cross section, light scattering by individual spheres was independent, and the scattering mean free path was estimated by $l_s = (\rho\sigma)^{-1}$ to be 500 µm, 100 µm, and 50 µm, respectively.

Lasing was achieved by optically exciting the dye solutions with a frequency-doubled Nd:YAG laser (wavelength $\lambda = 532$ nm) with 30 ps pulses at a repetition rate of 10 Hz. The pump beam was focused via a spherical lens onto the solution through the front window of the cuvette and the spot size was monitored by a charge-coupled-device (CCD) camera through the side window of the cuvette. The cuvette was rotated ~10° with respect to the pump beam to avoid feedback from the front window affecting the lasing modes [13].

In order to characterize the spatial coherence of the emission from these samples, we extended the technique utilized in [4] based on a Young's double slit experiment. The random laser emission exiting the front window of the cuvette (in the direction of the pump laser) was partially re-directed using a beam splitter. A spherical lens focused the random laser emission onto a screen with two slits, forming an image of the emission spot. The slits were 150 µm wide and separated by 750 µm. Behind the double slit, a CCD was positioned at the back focal plane of a cylindrical lens, oriented parallel to the slits, to measure the far-field interference pattern. The visibility of the interference pattern provided a measure of the coherence between pairs of points on the emission spot with a spatial separation equal to the double slit separation divided by the magnification of our imaging optics. Unless otherwise noted, we used a magnification of 6, thereby probing the spatial coherence between pairs of points separated by 125 µm. To ensure that the resolution of our imaging system did not artificially enhance the spatial coherence, we confirmed that the resolution of our imaging system (~15 µm) was significantly smaller than the double slit separation.

We first measured the spatial coherence of a weakly scattering sample ($l_s = 500$ µm) excited with a 215 µm diameter pump spot. Data corresponding to this sample/pump configuration is presented in the second column of Fig. 1. The single-shot emission spectra consisted of narrow lasing peaks on top of a broad-band amplified spontaneous emission (ASE). The interference fringe image is shown in the third row. The decay of the fringe visibility results from the temporal coherence and the finite slit width, so our discussion will focus on the visibility of the center fringe. The high contrast of the center fringe implied a high spatial coherence, which had not been observed before in a random laser. What was more surprising was that the fringes appeared in uniform vertical lines (parallel to the slits), and their position did not shift between pump pulses. This indicated that every pair of emission points imaged onto the two slits had the same phase difference, even for different pulses.

To confirm that the uniformity of the fringes was not an experimental artifact, we measured the interference pattern generated by a spatially coherent He:Ne laser ($\lambda$ = 632.8 nm, close to the random laser emission wavelength) scattered off the polystyrene spheres in the same sample. In this case, the phase of the scattered He:Ne laser light was scrambled and we observed rows of interference fringes, each with a random offset from the center of the two slits, similar to those presented in [4]. Because the phase difference between pairs of points incident on the double slit changed along the length of the slit, the fringes for different pairs of points appeared with varying offsets from the optical axis between the slits.

After eliminating the possibility of any artifact, we repeated the spatial coherence experiment on a sample of Rhodamine 640 in DEG without polystyrene spheres. In this sample we observed only ASE, as no scattering elements were present to provide feedback for lasing [14]. When we excited this sample with a similar-size pump spot, we again observed vertical fringes with high contrast. To explain this observation, we note that the excitation volume, imaged from the side, had a cone shape whose length was larger than the width. ASE was the strongest along the longest dimension of the gain volume because spontaneously emitted photons propagating in this direction experienced the most amplification. Since the excitation pulse was much shorter than the radiative decay lifetime of Rhodamine 640 molecules, most of the emission can be attributed to ASE originating from a few spontaneous emission events which were amplified along the axis of the excitation cone. As a result, the ASE leaving the front window of the cuvette had a uniform phase front, and generated vertical fringes. This behavior was similar to that of a superluminescent diode (SLD) which is known to exhibit high spatial coherence [15].

Our observation of spatially coherent ASE provided a clue for understanding the spatial coherence of laser emission from the weakly scattering sample. Although the emission from the weakly scattering sample consisted of both random lasing and ASE, the spatial coherence was nearly identical to the sample without scatterers, and thus we concluded that the random lasing component of the emission exhibited similar spatial coherence to the ASE component. To explain the high spatial coherence of the lasing component, we note that the scattering mean free path, $l_s$, was longer than the absorption length, $l_a$, of pump light, and the excitation volume had an elongated, conical shape similar to the sample without scatterers. The random lasing modes tend to orient themselves along the longest dimension of the gain volume. This mode orientation, combined with the weak scattering, allowed most of the lasing emission to leave the front cuvette window with a uniform phase front. Consequently, the interference fringes generated by each mode appeared at the same position. In agreement with this interpretation, we found that the spatial coherence from the same sample was reduced when we increased the pump area. As the pump spot diameter $d$ increased, the width of the excitation cone became comparable to the depth (the first row of Fig. 1), and light amplification along the cone axis was no longer stronger than in other directions. As the lasing modes reoriented themselves, the laser emission collected through the front cuvette window no longer exhibited a constant phase front. We also found that more modes were excited as we increased the excitation volume and eventually the peaks were so close to each other that they could no longer be resolved spectrally (the second row of Fig. 1). The presence of a large number of lasing modes with uncorrelated phase relationships reduced the spatial coherence of the laser emission (last row of Fig. 1).

Next, we switched to more strongly scattering samples and observed a decrease in the spatial coherence even for the smallest pump spot considered. The first column of Fig. 1 shows the data for a sample with $l_s$ = 50 μm and $d$ = 215 μm. As $l_s$ became smaller than $l_a$, the excitation volume changed from an elongated cone to a hemisphere. Meanwhile the number of lasing peaks increased for the same pump size because the stronger scattering reduced the effective volume of individual lasing modes. For the smallest pump spot size considered, there were already so many lasing peaks that they merged to form a continuous band in the emission spectra. Since the excitation volume was approximately hemispherical, there were no preferred directions for amplification and very low spatial coherence was observed.

To quantitatively describe the degree of spatial coherence, we computed the mutual coherence function, $\gamma$ from the interference fringe data. The degree of coherence between two fields, $E_1$ and $E_2$, is defined as $\gamma = \langle E_1 E_2^* \rangle / \sqrt{I_1 I_2}$, where $I_1 = |E_1|^2$, and $I_2 = |E_2|^2$. In our coherence measurement, the intensity on the two slits is equal and $\gamma$ reduces to the visibility: $\gamma = (I_{max} - I_{min})/(I_{max} + I_{min})$, where $I_{max}$ and $I_{min}$ are the maximum and minimum intensities of the interference fringes.

In Fig. 2, we compile $\gamma$ computed from measurements of four samples at six different pump spot sizes. In agreement with our qualitative discussion above, we found that the spatial coherence reduced monotonically with increasing pump area or decreasing scattering mean free path. A larger pump area or a shorter mean free path led to a more isotropic excitation volume and a greater number of lasing modes, both resulting in lower spatial coherence.

We also measured the coherence as a function of the spatial distance between points on the emission spot. Experimentally, the magnification of the imaging optics was changed so that pairs of points on the emission spot with varying separation were imaged onto the double slit. The magnification was adjusted by changing the distance between the cuvette and the spherical lens and the distance between the lens and the double slit. The data presented in the inset of Fig. 2 were taken from the sample with $l_s$=500 μm and $d$=215 μm. This corresponded to the elongated excitation volume shown in Fig. 2 and, as such, vertical interference fringes were visible at each magnification considered. Nonetheless, the coherence was seen to decrease with spatial distance. This observation is due to the finite volume of the individual lasing modes.

When the spatial coherence was probed at larger separations, the intensity of a single lasing mode was less likely to be equal at the two points. This amplitude imbalance led to the reduced fringe visibility.

Finally, we studied the effect of increasing the pump intensity on the spatial coherence of random laser emission, as shown in Fig. 3. We measured the interference patterns from 1× to 10× the lasing threshold, but did not observe a significant change in the visibility. Further, the lasing spectra continued to exhibit discrete peaks, even at 10× threshold. This observation agreed well with previous studies of weakly scattering random lasers in which spatial and spectral hole burning were found to limit the number of lasing modes, even well above threshold [16, 17]. Since the same lasing modes were excited as we increased the pump intensity, the degree of spatial coherence remained constant.

In summary, we have performed the first systematic, experimental investigation of the spatial coherence of laser emission from disordered media. We found that the spatial coherence of random lasers varies significantly with the scattering strength and the pump area. These observations were explained in terms of the number and characteristics of the random lasing modes, as dictated by the scattering length and the size and shape of the excitation volume. This work demonstrates the feasibility of utilizing random lasers as intense, spatially incoherent light sources for applications in which spatial cross-talk or speckle limits performance.

We thank A. Douglas Stone, Yidong Chong, and Christian Vanneste for stimulating discussions. This work is partially supported by NSF Grant No. DMR-0808937.

Fig. 1. Random laser characterization—each column corresponds to a sample/pump configuration defined by the scattering mean free path of the sample ($l_s$) and the pump spot diameter ($d$). Top row: Side-view images of the excitation volume (scale bar is 100 μm). Second row: Normalized emission spectra for a single pulse taken at 2× lasing threshold. Third row: Far-field interference fringe images (scale bar is 200 μm). Fourth row: Normalized, average cross-section of the interference fringes.

Fig. 2. Mutual coherence function $\gamma$ of laser emission from samples with varying mean free path $l_s$ (indicated in the legend) as a function of the pump spot diameter $d$. The inset shows $\gamma$ measured on the sample with $l_s = 500$ μm and $d = 215$ μm for various spatial distances between the pairs of points on the emission spot.

Fig. 3. The lasing spectra (left column) and interference fringes (right column) are presented at 3× (bottom row) and 8× (top row) lasing threshold for the sample with a mean free path $l_s = 500$ μm and excited with a pump spot of $d = 215$ μm.

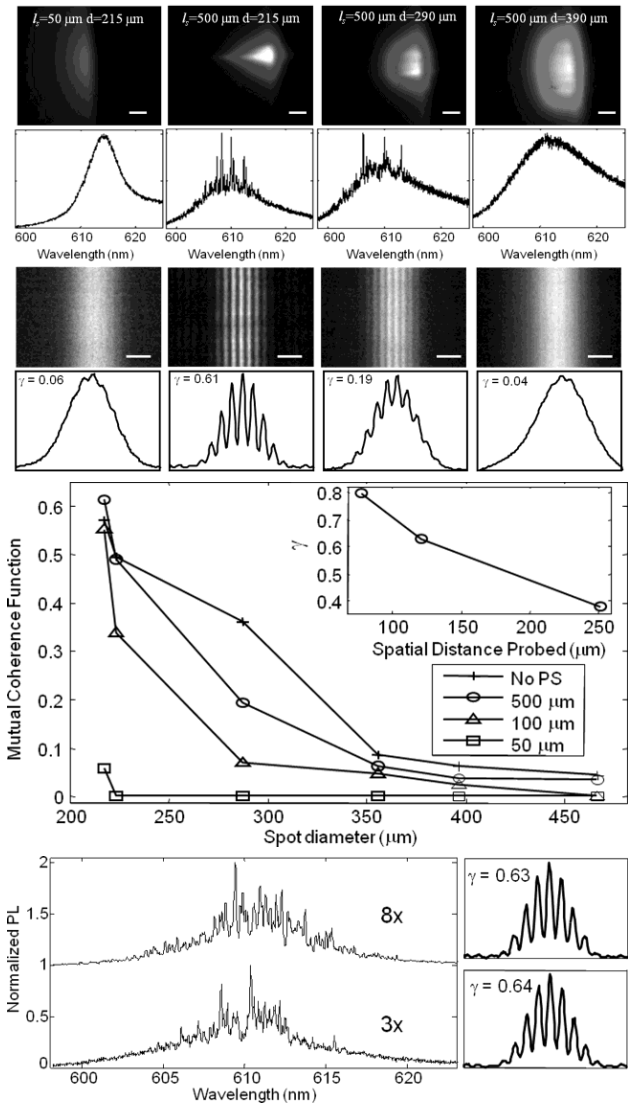